\documentclass[aip,jcp,texflow]{revtex4-1}
\usepackage{amsmath}
\usepackage[utf8]{inputenc}
\usepackage{xcolor}
\usepackage{amsfonts}
\usepackage{todonotes}
\usepackage[version=3]{mhchem}
\usepackage{bm}
\usepackage{multirow,tabularx}
\usepackage{braket}
\usepackage{siunitx}
\usepackage{xspace}
\usepackage[english]{babel}
\newcolumntype{K}[1]{>{\centering\arraybackslash}p{#1}}

\begin{document}

\title{Explicitly correlated formalism for second-order single-particle Green's function}
\author{Fabijan Pavo\v{s}evi\'c}
\affiliation{Department of Chemistry, Virginia Tech, Blacksburg, Virginia 24061, USA}

\author{Chong Peng}
\affiliation{Department of Chemistry, Virginia Tech, Blacksburg, Virginia 24061, USA}

\author{J. V. Ortiz}
\affiliation{Department of Chemistry and Biochemistry, Auburn University, Auburn, Alabama 36849-5312, USA}

\author{Edward F. Valeev}
\email{efv@vt.edu}
\affiliation{Department of Chemistry, Virginia Tech, Blacksburg, Virginia 24061, USA}

\begin{abstract}
We present an explicitly correlated formalism for the second-order single-particle Green's function method (GF2-F12) that does not assume the popular diagonal approximation,
and describes the energy dependence of the explicitly correlated terms. For small and medium organic molecules the basis set errors of ionization potentials of GF2-F12 are
radically improved relative to GF2: the performance of GF2-F12/aug-cc-pVDZ is better than that of GF2/aug-cc-pVQZ, at a significantly lower cost. 
\end{abstract}

\maketitle




The Green's function (GF), or propagator,\cite{linderberg2004propagators} formalism of many-body quantum mechanics is a complementary approach to traditional wave function methods for computation and interpretation of electronic structure. Whereas GF formalism is dominant in condensed phase physics as a step beyond the mean-field description, it has also enjoyed a sustained, albeit a less prominent, presence in the molecular electronic structure.\cite{ortiz2013electron,corzo2017chapter} The single-particle GF, or electron propagator, has primarily been employed as a computationally efficient route to post-mean-field ionization potentials (IP) and electron affinities (EA) and, more generally, spectral functions necessary to interpret various photoelectron spectroscopies; recently self-consistent GF theory has been revisited as a route to quantum embedding and to finite-temperature electronic structure.\cite{lan2017generalized,Phillips2014}

Here we present a general, {\em explicitly correlated} formalism for computing single-particle Green's functions. By employing many-body basis sets with explicit dependence
on the interelectronic distances, it is possible to formulate wave function methods with radically improved convergence to the analytic (complete basis set) limit.\cite{TenNo:2011bb,hattig2011explicitly,kong2011explicitly} In this work we demonstrate how
to achieve the same goal for the electron propagator. The initial validation of our approach focuses on computation of molecular ionization potentials with the second-order
approximation to the self-energy (GF2). The performance of the method is tested for ionization potentials (IP) on two sets of small to medium-sized molecules.\cite{corzo2015nr2,richard2016accurate,dolgounitcheva2016accurate}  
Our approach goes beyond the recent explicitly correlated correction to the GF2 ionization potentials of Ohnishi and Ten-no\cite{ohnishi2016explicitly} by incorporating
the energy dependence as well as extending beyond the diagonal approximation. Although the accuracy of the GF2 approximation is limited, it should be sufficient to gauge the effects of the energy dependence on the explicitly correlated contributions. Combination of our approach with higher-order nondiagonal Green's function approaches, such as the 2ph-TDA, ADC(3) and NR2 methods,\cite{von1984computational,ortiz1998nondiagonal} which are typically more robust than the second-order counterpart, is straightforward and will be reported elsewhere.

Ionization potentials (IPs) and electron affinities (EAs) are obtained from the poles of the electron propagator. With the zeroth-order defined as usual in ab-initio molecular electronic structure by the Hartree-Fock state, the poles are obtained from the Dyson equation:
\begin{align}
\label{eq:Dyson}
[\mathbf{F} + \mathbf{\Sigma}(E)]\mathbf{c} = E\mathbf{c},
\end{align}
where $\mathbf{F}$ is the Fock operator and $\mathbf{c}$ provides the Dyson orbitals. The energy-dependent self-energy operator $\mathbf{\Sigma}(E)$ incorporates the post-HF correlation and orbital relaxation effects. Within the superoperator formalism,\cite{goscinski1970moment,pickup1973direct} the self-energy operator is expressed as
\begin{align}
\label{SelfEnergy}
\mathbf{\Sigma}(E)=(\mathbf{a}|\hat{H}\mathbf{t})(\mathbf{t}|(E-\hat{H})\mathbf{t})^{-1}(\mathbf{t}|\hat{H}\mathbf{a}),
\end{align}  
where $\mathbf{a}\equiv\{a_{p}\}$ is a row vector of annihilation operators and $\mathbf{t}\equiv\{a_{p_1 p_2}^{p_3}, a_{p_1 p_2 p_3}^{p_4 p_5}, \dots \}$ contains the complementary two-, three-, and higher-body counterparts of $\mathbf{a}$. The orbitals occupied in the Hartree-Fock reference state are denoted by $i,j,\dots$ . The corresponding unoccupied orbitals expressed in
the orbital basis set (OBS) used to solve the Hartree-Fock equations are denoted by $a,b,\dots$. General OBS orbitals will be denoted by $p,q,\dots$, whereas their complement in the complete basis set (CBS) will be labeled by $\alpha,\beta,\dots$; the full set of CBS orbitals is denoted by $\kappa , \lambda, \dots$. Throughout this paper, we are using the standard tensor notation for products of annihilation ($a_p$) and creation ($a^p \equiv a^{\dagger}_{p}$) operators normal ordered with respect to the physical vacuum,\cite{kutzelnigg1997normal} i.e. $a^{\dagger}_{p_1} a^{\dagger}_{p_2} \dots a^\dagger_{p_n} a_{q_m} \dots a_{q_2} a_{q_1}=a^{p_1 p_2 \dots p_n}_{q_1 q_2 \dots q_m}$. In Eq. \eqref{SelfEnergy} the braket is defined as:
\begin{align}
(X|Y) \equiv \bra{0}[X^{\dagger},Y]_{+}\ket{0},
\end{align}
with
$|\hat{H}X) \equiv  [X,\hat{H}]_{-} \ket{0}$, where $\ket{0}$ is the Hartree-Fock reference wave function. The leading order correction to the self-energy occurs at the second-order perturbation theory obtained with the usual M\o ller-Plesset partitioning of the Hamiltonian:
\begin{align}
\label{eq:SE}
\Sigma_{pq}(E)&=(a_{p}|\hat{H}^{(1)}a_{kl}^{b})(a_{ij}^{a}|(E-\hat{H}^{(0)})a_{kl}^{b})^{-1}(a_{ij}^{a}|\hat{H}^{(1)}a_{q})\\\nonumber
&+(a_{p}|\hat{H}^{(1)}a_{cd}^{j})(a_{ab}^{i}|(E-\hat{H}^{(0)})a_{cd}^{j})^{-1}(a_{ab}^{i}|\hat{H}^{(1)}a_{q}),
\end{align}  
where the zeroth- and the first-order Hamiltonians are $\hat{H}^{(0)} \equiv \hat{F} = F^{\mu}_{\nu} a^{\nu}_{\mu}$ and $\hat{H}^{(1)}\equiv \hat{H} - \hat{H}^{(0)} \equiv \hat{W}$ ($\hat{F}$ and $\hat{W}$ are the Fock and fluctuation operators, respectively). Throughout this text, we are using the Einstein summation convention and real orbitals are assumed.

Evaluation of Eq. \ref{eq:SE} leads to
\begin{align}
\label{eq:SEfinal}
\Sigma_{pq}(E)&=\frac{1}{2} \sum_{ija} \frac{\bar{g}^{ij}_{pa}\bar{g}^{qa}_{ij}}{E+\epsilon_{a}-\epsilon_{i}-\epsilon_{j}}+\frac{1}{2}\sum_{iab}\frac{\bar{g}^{ab}_{pi}\bar{g}^{qi}_{ab}}{E+\epsilon_{i}-\epsilon_{a}-\epsilon_{b}}
\end{align}
where $\bar{g}^{pq}_{rs} \equiv g^{pq}_{rs} - g^{pq}_{sr}$ is the antisymmetrized Coulomb integral, where $g^{pq}_{rs} \equiv \bra{rs} r_{12}^{-1} \ket{pq}$. 
The first (2-hole-1-particle, or $2h1p$) term in Eq. \eqref{eq:SEfinal} describes the orbital relaxation effects and the second ($2p1h$) accounts for the electron correlation effects. As shown by Ten-no and Ohnishi,\cite{ohnishi2016explicitly} the sum over unoccupied states in the second term is slowly convergent in an atom. This can be seen immediately by recognizing that at $k$th zeroth-order poles the $2p1h$ contribution to the $k$th diagonal element of self-energy is a sum of the corresponding
MP2 pair energies:
\begin{align}
\label{eq:SEfinal2p1h}
\Sigma_{kk}^{2p1h}(\epsilon_k)&= \frac{1}{2} \sum_{iab} \frac{\bar{g}^{ab}_{ki}\bar{g}^{ki}_{ab}}{\epsilon_{k}+\epsilon_{i}-\epsilon_{a}-\epsilon_{b}} = \sum_{i} \epsilon^\text{MP2}_{ki}.
\end{align}
The slow basis set convergence is thus in direct analogy with
the slowly convergent error $\mathcal{O}[(L+1)^{-3}]$ of a truncated partial wave expansion of the atomic MP2 energy.\cite{kutzelnigg1992rates} Motivated by the close connection of the {\em diagonal} elements of the second-order self-energy at the corresponding Koopmans pole with the MP2 energy contributions, Ten-no and Ohnishi suggested an additive ({\em energy-independent}) explicitly correlated correction for the second-order self-energy using the explicitly correlated MP2-F12 pair energies. 

Here we demonstrate how to go beyond a simple additive correction by including proper {\em energy dependence} for the {\em nondiagonal} second-order Green's function method. The goal is to provide a robust reference for more approximate schemes such as that of Ten-no and Ohnishi, as well as establish an explicitly correlated approach for {\em general} GF methods. We start by augmenting the slowly convergent $\mathbf{t}=\{a^{i}_{ab}\}$ field operator with a geminal field operator. We postulate that the form of the geminal field operator is $\mathbf{t}^{\gamma}=\{\frac{1}{4}\tilde{R}_{ir}^{\alpha\beta}\tilde{a}_{\alpha\beta}^{i}\}$. The $\tilde{R}_{ir}^{\alpha\beta}$ is obtained from $R_{ir}^{\alpha\beta}$, the antisymmetrized matrix element of the geminal correlation factor $f(r_{12})$:\cite{tenno2004slater} 
\begin{align}
\label{eq:rmatelem}
R_{ir}^{\alpha\beta}\equiv\bra{ir}\hat{Q}f(r_{12})\ket{\alpha\beta}.
\end{align}
The tensors with tildes include the pair-spin projection due to singlet and triplet cusp conditions:\cite{tenno2004ratgen,zhang2012prediction} 
\begin{align}
\tilde{O}_{ij}^{\alpha\beta} &\equiv \frac{1}{2}(C_{0}+C_{1})O_{ij}^{\alpha\beta} + \frac{1}{2}(C_{0}-C_{1})O_{ji}^{\alpha\beta} \\\nonumber
&=\frac{3}{8}O_{ij}^{\alpha\beta} + \frac{1}{8}O_{ji}^{\alpha\beta}
\end{align}
where $C_{0,1}=1/2,1/4$ are the cusp coefficients for singlet and triplet pairs, respectively.\cite{kato1957eigenfunctions,Pack:1966fw}
Projector $\hat{Q}$ in Eq. \eqref{eq:rmatelem}
ensures that the geminal functions are orthogonal to Hartree-Fock as well as to the standard double excitations.

The explicitly correlated part of the self-energy is expressed as
\begin{align}
\label{SEF12}
\mathbf{\Sigma}(E)\leftarrow(\mathbf{a}|\hat{H}^{(1)}\mathbf{t}^{\gamma})(\mathbf{t}^{\gamma}|(E-\hat{H}^{(0)})\mathbf{t}^{\gamma})^{-1}(\mathbf{t}^{\gamma}|\hat{H}^{(1)}\mathbf{a}).
\end{align} 
Resolution of the matrix elements will give:
\begin{align}
(a_{p}|\hat{W}\mathbf{t}^{\gamma})=\frac{1}{4}\bar{g}^{\gamma\delta}_{kp}\tilde{R}^{kr}_{\gamma\delta}=\frac{1}{4}V^{kr}_{kp}
\end{align} 
while
\begin{align}
(\mathbf{t}^{\gamma}|\hat{W}a_{q})=\frac{1}{4}\tilde{R}_{is}^{\alpha\beta}\bar{g}_{\alpha\beta}^{iq}=\frac{1}{4}V^{iq}_{is}.
\end{align} 
The matrix elements of the resolvent are:
\begin{align}
(\mathbf{t}^{\gamma}|\hat{F}_N\mathbf{t}^{\gamma})&=\frac{1}{4}\delta^{i}_{k}\tilde{R}^{\alpha\beta}_{is}F^{\gamma}_{\alpha}\tilde{R}^{kr}_{\gamma\beta}-\frac{1}{8}F^{i}_{k}\tilde{R}^{\alpha\beta}_{is}\tilde{R}^{kr}_{\alpha\beta}\\
&=\frac{1}{4}\delta^{i}_{k}B_{is}^{kr}-\frac{1}{8}F^{i}_{k}X_{is}^{kr}
\end{align} 
and
\begin{align}
E(\mathbf{t}^{\gamma}|\mathbf{t}^{\gamma})=\frac{1}{8}E\delta^{i}_{k}\tilde{R}^{\alpha\beta}_{is}\tilde{R}^{kr}_{\alpha\beta}=\frac{1}{8}E\delta^{i}_{k}X_{is}^{kr},
\end{align}  
where $\hat{F}_N \equiv \hat{F} - E^{(0)}$ is the normal-ordered Fock operator.
The $V,X,B$ are standard F12 intermediates and their programmable expressions can be found elsewhere.\cite{pavovsevic2014geminal,pavosevic2016sparsemaps4,ChongCCSD2016,pavovsevic2017sparsemaps} Intermediates $V$ and $X$ were evaluated in the CABS approximation,\cite{valeev2004improving} while for intermediate $B$, we utilized approximation D.\cite{pavosevic2016sparsemaps4} 

This formalism has been implemented in a developmental version of Massively Parallel Quantum Chemistry package (MPQC) version 4.\cite{ChongCCSD2016} The implementation utilizes the TiledArray tensor library, which provides distributed parallel tensor routines, and hence the implementation is massively parallel.\cite{calvin2015scalable} Its performance is discussed in the next section.
 
We have assessed the performance of the new approach by computing the IPs of a set of 21 small molecules\cite{corzo2015nr2} and 24 medium sized organic electron accepting molecules (OAM24).\cite{richard2016accurate}
The basis set for the OBS that we used is aug-cc-pVXZ with the corresponding density-fitting basis set, aug-cc-pVXZ-RI, where X=D,T,Q,5 (X represents basis set cardinal number) and aug-cc-pVXZ-CABS basis set for the calculations with explicit correlations.\cite{dunning1989gaussian,kendall1992electron,woon1993gaussian,weigend2002ribasis,yousaf2009optimized} 
All computations were performed with the frozen-core approximation. The Slater-type correlation factor, $f(r_{12}) = (1-\exp(-\gamma r_{12}) / \gamma)$, with $\gamma=1.3, 1.9 $ and $2.1$ $\,\text{Bohr}^{-1}$ for the basis sets aug-cc-pVDZ, aug-cc-pVTZ and aug-cc-pVQZ has been used. 

In the following text, we denote the nondiagonal second-order Green's function method by GF2, whereas the GF2 method with fully energy-dependent explicitly correlated correction by GF2-F12. GF2(F12) denotes a hybrid approach in which the explicitly correlated correction is computed a posteriori, by solving the Dyson equation for the pole and Dyson orbitals at the GF2 level, followed by a single-shot evaluation of the explicitly correlated contribution to the self-energy with fixed energy and orbitals. Thus the difference between the GF2(F12) and GF2-F12 results will indicate the importance of the full energy dependence of the explicitly correlated terms in the second-order self-energy.

\begin{figure}[h!]
	\includegraphics[width=\columnwidth]{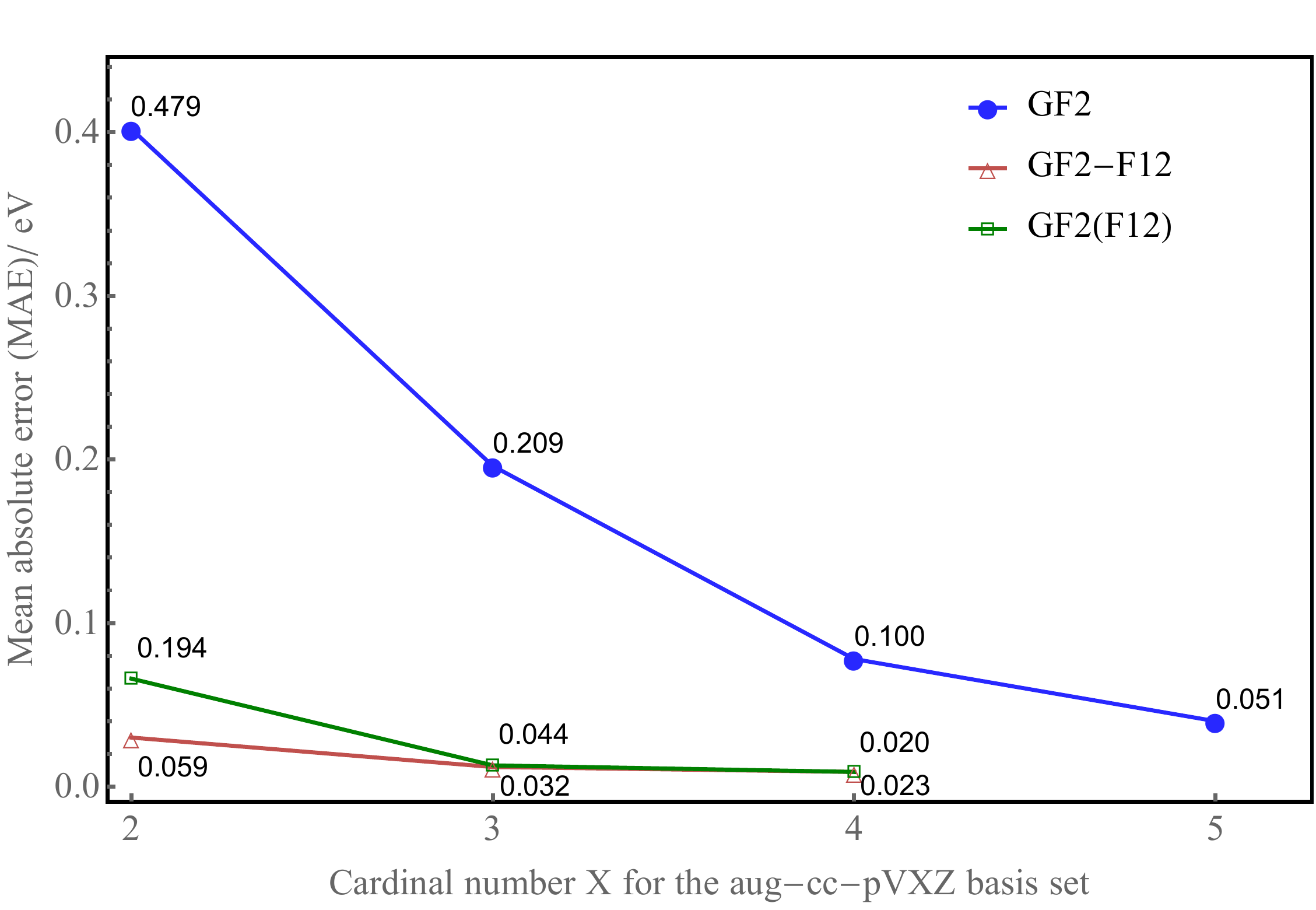}
	\caption{Mean absolute error in eV (MAE) for GF2 (blue), GF2(F12) (green) and GF2-F12 (red) methods with several basis sets. Values next to the points represent maximum absolute errors in eV (MaxAE).}
	\label{fig:MAEsmall}
\end{figure}

Table \ref{IPsmall} presents the statistical averages of the basis set errors of IPs for 21 small molecular systems obtained with the GF2, GF2(F12) and GF2-F12 methods. The complete basis set (CBS) limit of GF2 has been estimated using the two-point (aug-cc-pVQZ to aug-cc-pV5Z) $X^{-3}$ extrapolation scheme.\cite{halkier1998basis} Statistical analysis shows that the mean absolute errors in eV (MAE/eV) for GF2 are 0.402, 0.169, 0.078 and 0.040 with the basis sets aug-cc-pVXZ (X=D,T,Q and 5); GF2(F12) gives MAEs of 0.066, 0.013 and 0.009, while GF2-F12 produced MAEs of 0.030, 0.012 and 0.009 for the X=D,T and Q. The maximum absolute error (MaxAE/eV) in the case of the GF2 method is 0.479 (CH$_3$CH$_2$CH$_3$), 0.209 (HF), 0.100 (HF and HCl) and 0.051 (HCl) with respect to X=D,T,Q and 5; MaxAE with the GF2(F12) method is 0.194 (HF), 0.044 (HF) and 0.020 (HF), while in the case of GF2-F12 MaxAE is 0.059 (CH$_3$OH), 0.032 (CH$_3$F) and 0.023 (HF and CH$_3$F) for X=D,T and Q, respectively. These statistical parameters are shown graphically in Fig. \ref{fig:MAEsmall} where the x-axis is basis set cardinal number X and the y-axis represents the mean absolute error in eV (MAE). The values next to the points show MaxAE in eV. The IPs evaluated with the GF2(F12) and GF2-F12 approaches have dramatically smaller basis set errors than their GF2 counterparts. Furthermore, the rigorous GF2-F12 approach is preferred to the simpler GF2(F12) approach with the double-zeta basis.

Note that the use of approximation D for the B intermediate introduces negligible errors.\cite{pavosevic2016sparsemaps4} The maximum absolute error between D and the more rigorous approximation C\cite{kedvzuch2005alternative} does not exceed 0.010 eV in the case of the aug-cc-pVDZ basis set, while for the larger basis sets, the error vanishes completely. 

\begin{table}
	\centering
	\caption{Mean absolute error in eV (MAE) and maximum absolute error in eV (MaxAE) for the IPs of the data set of small 21 molecules with respect to the CBS limit for several basis sets.\label{IPsmall}}
	\begin{tabular} {  l | K{2cm} K{2cm}  K{2.1cm} | K{2cm} K{2cm} K{2.1cm} }\hline\hline
		& \multicolumn{3}{c}{MAE} & \multicolumn{3}{c}{MaxAE} \\
		Basis set & {GF2} & {GF2(F12)} & {GF2-F12} & {GF2} & {GF2(F12)} & {GF2-F12} \\ \hline
		aug-cc-pVDZ   & 0.402   & 0.066 & 0.030 & 0.479$^a$ & 0.194$^b$ & 0.059$^c$\\ 
		aug-cc-pVTZ   & 0.169  & 0.013 & 0.012 & 0.209$^b$ &  0.044$^b$ & 0.032$^d$\\ 
		aug-cc-pVQZ   & 0.078  & 0.009 & 0.009 & 0.100$^e$ & 0.020$^b$ & 0.023$^f$\\ 
		aug-cc-pV5Z   & 0.040  & - & - & 0.051$^e$ & - & - \\ 
		\hline\hline		
	\end{tabular}
	\\$^a$CH$_{3}$CH$_{2}$CH$_{3}$, $^b$HF,  $^c$CH$_{3}$OH, $^d$CH$_{3}$F, $^e$HF and HCl, $^f$HF and CH$_{3}$F\\
	
\end{table}

We also tested the explicitly correlated GF2 approaches by computing IPs of medium sized molecules in the OAM24 data set. The IP basis set error statistics are given in the Table \ref{IPOAM24}. The CBS limit has been calculated with the GF2 method using the two-point extrapolation scheme (aug-cc-pVTZ to aug-cc-pVQZ).\cite{halkier1998basis}

\begin{table}
	\centering
	\caption{Mean absolute error in eV (MAE) and maximum absolute error in eV (MaxAE) for the IPs of the OAM24 data set with respect to the CBS limit for several basis sets.\label{IPOAM24}}
	\begin{tabular} {  l | K{2cm} K{2cm}  K{2.1cm} | K{2cm} K{2cm} K{2.1cm} }\hline\hline
		& \multicolumn{3}{c}{MAE} & \multicolumn{3}{c}{MaxAE} \\
		Basis set & {GF2} & {GF2(F12)} & {GF2-F12} & {GF2} & {GF2(F12)} & {GF2-F12} \\ \hline
		aug-cc-pVDZ   & 0.372  & 0.081 & 0.013 & 0.459$^a$ & 0.145$^b$ & 0.039$^c$\\ 
		aug-cc-pVTZ   & 0.151  & 0.021 & 0.007 & 0.183$^a$ &  0.054$^d$ & 0.016$^e$\\ 
		aug-cc-pVQZ   & 0.064  & - & - & 0.077$^a$ & - & -\\ 
		\hline\hline		
	\end{tabular}
	\\$^a$maleic anhydride, $^b$naphthalenedione, $^b$nitrobenzene, $^c$Cl$_{4}$-benzoquinone, $^d$benzoquinone, $^e$phenazine\\
	
\end{table}

The results from Table \ref{IPOAM24} are in agreement with those presented for the small molecules and support the same conclusion. GF2-F12 shows very small errors, giving a MaxAE of only 0.039 eV in the case of the aug-cc-pVDZ and making this method much more accurate than the more costly alternative provided by GF2 with the aug-cc-pVQZ basis. GF2-F12 is also preferred to the more approximate GF2(F12) approach, which suggests the explicitly correlated correction to self-energy should
indeed include the energy dependence properly.

We have presented and efficient, massively parallel implementation of the explicitly correlated nondiagonal energy dependent GF2-F12 method. By including geminal field operators, we account for the missing electron correlation effects due to the incompleteness of the basis set. Numerical tests on small and medium molecules suggested that the ionization potentials computed with the new explicitly correlated GF2-F12 method in conjunction with the modest aug-cc-pVDZ basis set had smaller basis set errors than their non-explicitly-correlated GF2 counterparts with a much larger aug-cc-pVQZ basis set. The computation time of the first ionization potential of pyridine with the GF2/aug-cc-pVQZ method takes 404 while GF2-F12/aug-cc-pVDZ method takes only 91 seconds (using four computing nodes with 24 Intel Xeon E5-2680 v3 2.50 GHz CPU cores). The corresponding absolute basis set errors are 0.061 and 0.018 eV, respectively. The proposed GF2-F12 method is easily extensible to higher-order Green's function approaches; such efforts will be reported elsewhere. 

All calculated ionization potentials for GF2, GF2(F12) and GF2-F12 methods can be found in the supplementary material [URL will be inserted by AIP].

We would like to thank Mr. Cannada A. Lewis for useful discussions. FP, CP and EFV acknowledge the support by the U.S. National Science Foundation (awards CHE-1362655 and ACI-1450262). JVO acknowledges the support of the U.S. National Science Foundation via grant CHE-1565760 to Auburn University. We also acknowledge the computational resources of Advance Research Computing at Virginia Tech (www.arc.vt.edu).

\bibliography{refs}

\pagestyle{empty}

\end{document}